\documentstyle[12pt,aasms4]{article}

\def\etal{{\it et al.~}} 
\def\deg{\ifmmode^\circ _\cdot\else$^\circ _ \cdot$\fi }    
\def\degg{\ifmmode^\circ \else$^\circ $\fi } 
\def\arcs{\ifmmode {'' }\else $'' $\fi}     
\def\arcm{\ifmmode {' }\else $' $\fi}     
\def\buildrel#1\over#2{\mathrel{\mathop{\null#2}\limits^{#1}}}
\def\mper{\ifmmode \buildrel m\over . \else $\buildrel m\over .$\fi}
\def\hper{\ifmmode \rlap.^{h}\else $\rlap{.}^h$\fi}
\def\sper{\ifmmode \rlap.^{s}\else $\rlap{.}^s$\fi}
\def\arcsper{\ifmmode \rlap.{' }\else $\rlap{.}' $\fi}
\def\arcmper{\ifmmode \rlap.{'' }\else $\rlap{.}'' $\fi}

\def\et{{\it et~al.~}}

\newcommand{\ea}{{\it et~al.\/} }
\newcommand{\cov}{{\bf V} }
\newcommand{\sa}{\sigma_{\alpha}}
\newcommand{\sd}{\sigma_{\delta}}
\newcommand{\sct}{\sigma_{c}}
\newcommand{\slt}{\sigma_{l}}
\newcommand{\tij}{\theta_{ij}}
\newcommand{\cm}{C_{M}}
\newcommand{\lla}{\bar{\ell}}    

\def\apj{ApJ}  
\def\apjl{Ap.~J.~ } 
\def\aasup{AAS }     

\lefthead{Femen\'{\i}a \etal}
\righthead{IAC-Bartol 1994 campaign results}

\begin{document}

\title{The IAC-Bartol Cosmic Microwave Background Anisotropy Experiment: Results of the 1994 Campaign}

\author{B. Femen\'{\i}a\altaffilmark{1,3}, R. Rebolo\altaffilmark{1,4},
 C.M. Guti\'errez\altaffilmark{1,5}, M. Limon\altaffilmark{2,6}, L. Piccirillo\altaffilmark{2,7}}

\altaffiltext{1}{Instituto de Astrof\'\i sica de Canarias, 38200 La Laguna,
Spain}

\altaffiltext{2}{Bartol Research Institute, University of Delaware, Newark, DE 
19716}

\altaffiltext{3}{bfemenia@ll.iac.es}

\altaffiltext{4}{rrl@ll.iac.es}

\altaffiltext{5}{cgc@ll.iac.es}

\altaffiltext{6}{limon@pupgg.princeton.edu}

\altaffiltext{7}{picciril@wisp5.physics.wisc.edu}

\begin{abstract} 

	We present the results of a Cosmic Microwave Background (CMB) anisotropy ground-based
millimetric experiment sensitive to fluctuations on angular scales of $\sim 2\degg$.  Four
independent bands centered at 3.3, 2.1, 1.3 and 1.1 mm collected $\sim 550$ hours of observation during
the Summer of 1994. The instrument was located on the island of Tenerife at an altitude of 2400
m. The low water-vapor content and the atmospheric stability of the site, combined with new
techniques to subtract  atmospheric noise, result in the reduction of atmospheric contamination
in the lowest frequency channel to a level of $\sim 1.5$~times the instrument noise. Detailed
estimations of  Galactic foreground contamination show that this contribution is negligible at $|b|
\stackrel{>}{_\sim} 12\degg$. Two different multipole bands ($\ell=53^{+22}_{-13}$ and
$33^{+24}_{-13}$) are analyzed showing that our technique to subtract the atmospheric contribution
is more effective in the multipole band at $\ell=53$. A likelihood analysis of these data reveals the presence of a
common signal between the channels at 3.3, 2.1 and 1.3 mm corresponding to a band power estimate of
$ \sqrt{\lla (\lla+1) C_{\lla}/(2 \pi)} \, = \, 2.0^{+1.0}_{-0.8} \cdot 10^{-5}$ and $\sqrt{\lla
(\lla+1) C_{\lla}/(2 \pi)} \, = \, 4.1^{+2.4}_{-2.2} \cdot 10^{-5}$ for the $\ell =53$ and 33
multipole bands respectively. Calibration uncertainty has been treated as a systematic effect. The
level of fluctuations in the $\ell =53$ band is in good agreement with our preliminary analysis
presented in Piccirillo \et 1997, with measurements by other experiments working at similar angular
scales, and with the predictions of standard Cold Dark Matter (CDM) models.

\end{abstract}


\section{INTRODUCTION}
\label{intro}

	In recent years the COBE DMR (\cite{ben96}), FIRS (\cite{firs}) and Tenerife (\cite{ten94})
experiments have successfully detected anisotropy at large angular scales yielding precise
estimations of the overall normalization of the power spectrum (\cite{hanc97a}) and its shape at low
multipole moments $\ell \stackrel{<}{_\sim} 30$. At smaller angular scales ($\ell
\stackrel{>}{_\sim} 70$) a wealth of experiments have reported detections of anisotropy (for an
up-to-date list of experimental results see Tegmark's web site:
http://www.sns.ias.edu/$\sim$max/cmb/experiments.html). These detections clearly suggest a increase
in the power spectrum at $\ell \sim 200$ (\cite{sk95},\cite{hanc97b}) and its decline at $\ell \sim
400$ (\cite{cat2}). The main aim of our experiment (for preliminary results see \cite{lucio3}) is to
fill the gap in $\ell$-space remaining between the two $\ell$-space regions described above. Thus,
with our observing strategy we are probing the multipole bands at $\ell = 33^{+24}_{-13}$ and $\ell
= 53^{+22}_{-13}$, in between the Tenerife experiment at $\ell = 20 \pm 8$ (\cite{hanc97a}) and the
ACME South Pole at $\ell = 68^{+38}_{-32}$ (\cite{sp94}).

	The choice of the observing frequencies also distinguishes this experiment from other
ground-based experiments. It is known that the combined contributions from Galactic foregrounds and
discrete radio sources reaches its minimum at about 55 GHz. However, the presence of a strong
$O_{2}$ absorption line in the range 50-70 GHz, and the increasing atmospheric emission at
millimetric wavelengths hamper observations at these frequencies forces ground-based experiments to
observe at cm wavelengths where contamination from synchrotron and free-free emission are of
concern. Thus, the millimetric range has been traditionally exploited by satellite and balloon
experiments, and only very few ground-based experiments have observed at these frequencies (Python
(\cite{pythonIII}), SuZIE (\cite{suzie}) and \cite{andr91}).

	In section~\ref{instrument} we briefly describe the instrument and  concentrate on the
measurement technique. In section~\ref{observations} we give the details of the
observations. Atmospheric effects in our data are described in section~\ref{atmosphere}. The
details of the atmospheric noise reduction technique and the resulting final data sets are presented
in section~\ref{technique}. In section~\ref{foregrounds} we conclude that at high $|b|$ the only
expected signal is the CMB. Section~\ref{statistics} describes the statistical data, and conclusions are
presented in section~\ref{conclusions}.

\section {INSTRUMENTAL SET-UP AND MEASUREMENT TECHNIQUE}
\label{instrument}

	The instrument has been described in detail in previous work (\cite{lucio1};
\cite{lucio2}). The optics forms an off-axis Gregorian telescope with a parabolic primary mirror (45
cm diameter) and a hyperbolic secondary mirror (28 cm diameter). The detector is a four-channel
photometer equipped with $^{3}$He bolometers working at 0.33K. The bands peak at 3.3, 2.1, 1.3 and
1.1 mm wavelengths (channels 1, 2, 3 and 4 respectively). The beam response for all channels can
be well approximated by a Gaussian with FWHM=$2.^{\circ}03 \, \pm \, 0.^{\circ}09$ and no
significant sidelobes are found.

	In addition to the optics, the response of the instrument to a point-like source depends on
the observing strategy and demodulation of the data (\cite{ws95}). The observing strategy consists
in daily drift scans at constant declination achieved by fixing the telescope in azimuth ($\phi$)
and elevation ($\theta$). Additionally the beam moves on the sky as the primary mirror wobbles
sinusoidally while the secondary is fixed.  The right ascension ($\alpha$) and declination
($\delta$) at which the antenna is pointing at time $t$ are given in a good approximation by:
$\delta(t) = \delta_{0}$ and $ \alpha(t)\, =\, \alpha_{0}\, +\, \beta_{0}/\cos (\delta_{0}) \times
\sin(2 \pi f_{w}t+ \epsilon)$, where $(\alpha_0,\delta_0)$ is the initial position of the antenna,
$\beta_{0} \,=\, 2.6\degg$ is the zero-to-peak chopping amplitude at a reference frequency $f_{w}\,
=\, 4$ Hz and $\epsilon$ is an irrelevant phase constant. Each bolometer's output is sampled at
$f_{s} \, = \, 80$ Hz coherently with the mirror movement so to have 20 samples per mirror
oscillation. The signal is demodulated in software by evaluating the amplitude of the first (4 Hz)
and second (8 Hz) harmonic of the reference frequency $f_{w}$. Each demodulation (1F and 2F for the
first and second harmonic respectively) can be divided in two components: the in-phase component
containing mostly the sky-signal coherent with the reference motion plus the instrument noise and the
out-phase containing mostly the instrument noise plus other sources of systematic noise. Such division
requires a careful choice of phase which was obtained from observation of the Moon transiting the
instrument beam. Then the mapping function at a sky location of coordinates $(\alpha,\delta)$ when
the antenna is pointing towards $(\alpha_{0},\delta_{0})$ is computed as (\cite{ws95}):

\begin{equation}
M_{nF}(\alpha,\delta;\alpha_{0},\delta_{0}) \, =  \,
\frac{N_{n}}{\sqrt{2}\pi \sigma^{2}} \, f_{w} \, \int_{-1/(2f_{w})}^{1/(2f_{w})} dt \, \cos (n \, 2 \pi f_{w} t \, + \zeta) \, \exp \left[ - \frac{\Delta^{2}(t)}{2 \sigma^{2}} \right]
\end{equation}

\noindent for $n=1,2$ indicating respectively the 1F and 2F demodulation, $\sigma \, = \,
FWHM/\sqrt{8 \ln 2}$, $\zeta$ is a phase constant, $\Delta(t)$ is the angular distance between the
point of coordinates $(\alpha,\delta)$ and the center of the beam at time $t$ and $N_{n}$ is a
normalization constant for the $n$ demodulation obtained by requiring an output of $1$~K when the
input is an extended source of $1$~K filling the positive lobe. 

	The response of the 1F and 2F demodulations resembles the usual 2-beam and 3-beam responses
to the transit of a point-like source through the  beam. The 1F demodulated data are
well fitted by a 2-beam response with asymmetric Gaussians with $\sigma_{\alpha} \, = \,
1.^{\circ}03$ in the RA direction and $\sigma_{\delta} \, = \, 0.^{\circ}86$ in the declination
direction, and with a beam throw of $\beta_{0} \, = \, 2.^{\circ}38$. The 2F demodulated data are
fitted by a 3-beam response with $\sigma_{\alpha} \, = \, 1.^{\circ}56$, $\sigma_{\delta} \, =
\,0.^{\circ}86$ for the positive lobe, and $\sigma_{\alpha} \, = \, 0.^{\circ}89$, $\sigma_{\delta}
\, = \,0.^{\circ}86$ for the negative lobes and a beam throw of $\beta_{0} \, = \, 2.^{\circ}40$.
Additionally, these fits must be multiplied by the factors $\aleph_{1F} \, = \,0.362 $ and
$\aleph_{2F} \, = \, 0.593$ to yield the normalizations to the actual response functions for 1F and
2F respectively. These approximations greatly simplify the statistical analyses as discussed in
section~\ref{statistics}.

	The response of the instrument to different angular scales (represented in terms of the
 corresponding multipole moment $\ell$) is completely specified by the window function. For a
 constant declination observation the window function for the product of two temperatures separated
 by an angle $\psi$ is computed according to \cite{ws95} to give:

\begin{equation}
W_{\ell}(\psi)_{nF} \, = \, N_{n}^{2} \, B_{\ell}^{2}(\sigma) \, \sum_{r=0}^{\ell} \,
\frac{(2\ell-2r)!(2r)!}{[2^{\ell}r!(\ell-r)!]^2} \, J^{2}_{n}[(\ell - 2r) \beta_{0}] \, \cos[(\ell-2r)\psi] \,
j^{2}_{0}\left[\frac{(\ell-2r) \Delta \Phi}{2}\right]
\end{equation}

\noindent where $B_{\ell}$ refers to the beam profile: $B_{\ell}(\sigma) \, = \, \exp [-\ell (\ell
+1) \sigma/2]$, $J_{n}$ is the n-th order Bessel function of the first kind, $j_{0}$ is the
zeroth-order spherical Bessel function and $\Delta\Phi$ is the bin size in radians on the sky. In
figure~\ref{fig_window} we show the window functions for $\psi \, = \, 0$ and $\Delta \Phi = 4$~min
in RA, corresponding to the $\ell$-ranges $\ell_{1F} \, = \, 33^{+22}_{-13}$ and $\ell_{2F} \, = \,
53^{+22}_{-15}$. The central value corresponds to the band power average and the upper and lower
limits give the $\ell$ ranges where $W_{\ell}(0)_{n}$ has amplitudes larger than $e^{-1/2}$ times
its peak.

\section{CALIBRATION AND OBSERVATIONS}
\label{observations}

	Laboratory calibrations were performed by placing blackbody radiators at different
temperatures in front of the optical window. By means of an off-axis mirror the optics within the
cryostat is redirected towards a vessel containing eccosorb and divided in four sections. Two
sections were filled with liquid Nitrogen alternately placed between the other two sections which
were filled with liquid Oxygen.  A measurement of the pressure at which these two liquids evaporate
gives a precise measurement of the temperature. Then the container is rotated at 2 revolutions per
second so the detectors see two black bodies of known temperatures. The calibration factors thus
obtained have uncertainties ranging from 1.5\% (Channel 1) up to 7.6\% (Channel 4). Since
we are concerned with the possibility of systematic effects in the calibration process we quote a
laboratory calibration uncertainty  based on our observations of the Moon (see below).

	Our observations were conducted during  June and July 1994, collecting about 550 hours of data, at
Observatorio del Teide at Tenerife (Spain). The site, at an altitude of 2400 m and latitude N
$28^{\circ}29$, is well known among the solar community for its good seeing and the large percentage
of clear days, the latter due to the fact that the inversion layer lies below the observatory for
about 75 \% of the time. This is also the location of the Tenerife experiment (\cite{ten94}) which
has already demonstrated the potential of the site for hosting observations in the cm range due to
the  stability of the atmosphere and its excellent transparency (\cite{Dav96}). These features
makes of this site a promising place for mm observations.

	The observations concentrated at declination $\delta = 40^\circ$ by pointing the antenna
towards the North meridian and at an elevation angle of $h = 78^{\circ}.7$. This declination has been
extensively measured from this site at larger angular scales($\sim 5^\circ$) and lower frequencies
(10, 15 and 33 GHz) with reported detections of structures in the CMB by the Jodrell Bank-IAC
experiments (\cite{ten94};\cite{gutierrez2}).

	 The measurements started after the Sun was well below the horizon in order to avoid solar contamination. The
angular distance between Moon and beam was always greater than $23^{\circ}.5$ for the $\delta = 40^\circ$ observations.
In addition the Moon was observed for astronomical calibration.  The values obtained from the analysis of the Moon
transits in the 1F demodulation agree within $\sim~20~\% $ of the laboratory calibrations for all channels (see
figure~\ref{fig_moon}). The bulk of the calibration uncertainty (6~\%, 26~\%, 22~\% and 10~\% for channels 1, 2, 3 and 4
respectively) we believe is due to the lunar models used (\cite{moon_cal,hagfors}). By fitting the observed Moon transit
to the expected transits at both demodulations, and modeling the Moon as an extended uniform disk and the beam as
Gaussian shaped, we obtain estimates for the beam width, $\sigma$, and chopping amplitude, $\beta_{0}$, given in
table~\ref{tab_fits_moon}. Since the Moon transit is better defined in the 1F data we adopt the estimates for $\sigma$
and $\beta_{0}$ from the 1F demodulation. The values for $\beta_{0}$ from the fits to each channel are consistent with
each other, yielding an average value of $\beta_{0}= 2^{\circ}.90 \pm 0^{\circ}.03$. As expected, the beam widths seen by
each channel show a monotonic decrease with frequency so that channel 1 has the widest beam and channel 4 the narrowest
beam. Even so, given the sizes of the error bars assigned to each width, all of them are consistent with a unique beam
width given by the weighted average: $\overline{\sigma}= 0^{\circ}.86 \pm 0^{\circ}.04$. In any case, for calibration
purposes we adopted for each channel the best fit value obtained for that channel. It is worth mentioning the agreement
between the FWHM estimated for channel 1 in the lab ($FWHM = 2^{\circ}.40 \pm 0^{\circ}.10$) with that from the fit to
the Moon transit ($FWHM = 2^{\circ}.21 \pm 0^{\circ}.24$ from the 1F Moon transit and $FWHM = 2^{\circ}.4 \pm
0^{\circ}.6$ from the 2F Moon transit). The discrepancy between the values of $\beta_{0}$ measured in the lab and from
the Moon transit can only be attributed to the antenna assembling, and throughout this work we have assumed the value of
$\beta_{0}$ as obtained from the Moon transit analysis.

\section{ATMOSPHERIC EFFECTS ON THE DATA}
\label{atmosphere}

	The biggest source of noise for CMB ground observations at millimetric wavelengths comes
from atmospheric emissions dominated by fluctuations of $O_{2}$ and water vapor contents. Given the
 stability of the pressure during the campaign, the most important parameter turned out to be
the precipitable water vapor. The meteorological conditions during the campaign are summarized in
figure~\ref{fig_weather}. The pressure and temperature are measured at the observatory four times a
day. The relative humidity is also measured at the observatory every few minutes. The precipitable
water vapor is obtained from the measurements of the balloons launched twice a day by the Spanish
Meteorological Institute from sea level. In our case, the precipitable water vapor is obtained by
integrating the balloon measurements from the Observatory level up to $ 12 - 15$ Km.

\subsection{ATMOSPHERIC NOISE LEVELS VS INSTRUMENT NOISE LEVELS}

	Atmospheric effects in our data become evident as a big increase with respect to the
expected instrument noise levels.  In the ideal case (i.e. the instrument noise is white) we should
observe flat power spectra if our data were entirely due to instrumental noise. However our data are
a combination of instrument noise, atmospheric noise and astronomical signal, the latter to be
ignored given its weakness in each night of observation. At low frequencies neither the instrument
noise nor the atmospheric noise exhibit flat spectra. Bolometers show a characteristic $1/f$
spectrum while atmospheric fluctuations have a more complicated spectrum and in general follow a
$1/f^{k}$ spectrum with $k$ variable. On the other hand, both spectra flatten at high frequencies,
becoming indistinguishable from each other. At high frequencies the  power spectra are flat,
and assigning to the instrument noise the value corresponding to such noise floor is a huge
overestimation.

        In theory, this problem could be solved by using the out-phase component produced during the
demodulation because it is expected to contain only instrumental noise. In practice, even in the
out-phase component there are still residual amounts of atmospheric noise because it is impossible
to find  a constant demodulation phase which completely sets to zero the sky-signal in the
out-phase component. In figure~\ref{fig_ps1} we show the noise spectra in thermodynamic temperature
for all channels and both demodulations of the in- and out-phase components for a typical night of
observation. Channel 1 exhibits power spectra almost flat in the out-phase spectra for 1F and 2F
demodulations, indicating that most of the noise in channel 1 at high frequencies is due to
instrument noise. The out-phase components for the rest of the channels still contains considerable
amounts of atmospheric noise as indicated by the similar shapes of the spectra in the in- and
out-phase components.  This interpretation is strongly supported by the fact that the spectral shape at low frequencies,
and the noise floor  at high frequencies,  change from day to day.  The higher values for the instrumental noise as obtained from
the analysis of the out-phase components of the the 1F data indicate the greater ability of the 2F
demodulation in removing linear gradients caused by atmospheric emission .  The final upper limits
in thermodynamic units assigned to the instrument noise are 2.8, 0.8, 1.5 and 1.0 mK~s$^{1/2}$ for
channels 1 to 4 as obtained from the analysis of the out-phase components of the 2F demodulation. In
figure~\ref{fig_histo1} we show the distribution of the $rms$ values, calculated over 30 minute
intervals, with 10 second bins for all channels and both demodulations.  At 10 seconds the
contribution from instrument noise to the $rms$ would be less than 1mK for all channels, so most of
the noise must be atmospheric.  The  noise level and the width of its distribution are
larger in the 1F demodulation. Additionally, the minimum values obtained in the 2F data are smaller; another
indication of the enhanced ability of the 2F demodulation to reduce atmospheric noise.

\subsection{ATMOSPHERIC CORRELATIONS IN THE DATA}

	The second relevant effect due to atmospheric contamination concerns correlation between
different channels. This is evident from the highly correlated time-variable signals seen by all
channels for the same night of observation. This point will be used subsequently on as the cornerstone for our analysis to
reduce atmospheric noise in the data. In table~\ref{tab_correlations1} we present the mean
correlation between channel $i$ ($i=1,2,3$) and channel 4 for the whole campaign and for the data
selected to build the final data sets. As expected this correlation increases when increasing
the frequency of channel $i$ and it is higher in the 1F data than in the 2F data.

	Another effect is the auto-correlation introduced by atmospheric noise. Primary evidence of
this effect is that the $rms$ of the data  do not follow the $1/\sqrt{N}$ law when the data are binned. We
have computed the auto-correlation function for the data within the same night of observation as seen
by each channel at both demodulations. This was done by using the power spectra of the in-phase
components and the Wiener-Khinchin relations (see e.g. \cite{barkat}). As noticed when computing the
instrument noise limits, the in-phase component power spectra show a large departure from white
noise, implying non-zero auto-correlations at lags different from zero. The average auto-correlation
function for the whole campaign and for all channels and both demodulations is shown in
figure~\ref{fig_auto-correlation}.  For a given lag we observe that the auto-correlation function is
always larger in the high frequency channels.  The coherence time is always larger for the 1F data
than for the 2F data. This reflects the higher efficiency with which a double-switching technique is
able to discard the effects of linear temporal drifts due to atmospheric emission. From
figure~\ref{fig_auto-correlation} we conclude that it is necessary to use bins large enough such that
correlation between adjacent bins is reduced (see section~\ref{technique}). This will ensure that the standard deviations are
properly computed and assigned as error bars when moving to larger bin sizes.

\section{DATA PROCESSING}
\label{technique}

	The demodulation process produces a data point every 0.25 s per each 1F and 2F demodulation
and per channel. During the demodulation 3.3\% of the data was rejected due to problems of
synchronism between the mirror movement and the data acquisition system. The bulk of the data
rejection was performed during the subsequent phases of binning and editing comprising the data
processing. After demodulation a binning is performed to bring the data from 0.25 s to 10 s. This is
a trade-off between the need to reduce correlation between adjacent bins (auto-correlation $<65 \%$
for all channels at 1F demodulation and $<35 \%$ for all channels at 2F demodulation) and to have a
large enough number of points for the linear fits and further binning processes involved in the
cleaning technique. During the 0.25 s to 10 s binning an iterative 3$\sigma$ filter is applied three
times to discard glitches due to malfunctions of the data acquisition system. Bins of 10 s built
with 15 or less points at 0.25 s are also discarded. In total, the amount of data rejected in this
binning is about  8\% for all channels and both demodulations.

	As explained in the section~\ref{atmosphere}, the extremely large temporal correlation
between all channels and the amplitudes and behavior of the noise levels with the bin size strongly
support the atmospheric origin of the bulk of the noise in our data. Hence, the primary goal of any
data reduction concerns the assessment and subtraction/reduction of this unwanted source of noise.

\subsection{DESCRIPTION OF THE CLEANING TECHNIQUE}

	The approach adopted in this work consists in exploiting the high correlation between
channels as an indicator of the atmospheric emission, so that by using a channel as a monitor we can
clean the rest of the channels. Since channel 4 (1.1 mm band) is the most sensitive to atmospheric
emission, we adopt it as the monitor. The method assumes that at each channel $i$ we have a
superposition of astronomical signal attenuated by $f_{i}$ due to the atmospheric transparency, plus
the contribution from the atmospheric emission. When expressing all quantities in antenna
temperature we have at channel $i$: $\Delta T_{ANT,i}=f_{i} \Delta T^{astro}_{ANT,i}+\Delta
T_{ANT,i}^{atm} $. We assume that the atmospheric contributions are perfectly correlated between
different channels so: $\Delta T_{ANT,i}^{atm}= \alpha_{i}\,\Delta T_{ANT,4}^{atm}$. This assumption
is strongly justified given the high correlation between data at different channels taken during the
same night of observation (see table~\ref{tab_correlations1}). The amplitude of the signals are too
high to be attributed to CMB signal or other astronomical signal (see section~\ref{foregrounds})
giving additional support to the above assumption. To recover the astronomical signal at channel $i$
in thermodynamic temperature ($\Delta T^{astro}_{i}$) and referred to the top of the atmosphere we
solve the linear equation:

\begin{equation} 
\label{eq_main}
\Delta T_{ANT,i} =\frac{f_{i}}{c_{i}} \Delta T^{astro}_{i}+(\Delta
T_{ANT,4}-\frac{f_{4}}{c_{4}}  \frac{1}{\rho_{i4}} \Delta T^{astro}_{i}) \times
\alpha_{i} 
\end{equation}

\noindent where $\Delta T_{ANT,i}$ and $\Delta T_{ANT,4}$ are the recorded antenna temperatures at
channels $i$ and 4 respectively; $f_{i}$ and $f_{4}$ the atmospheric attenuation factors; $c_{i}$
and $c_{4}$ are the Rayleigh-Jeans to thermodynamic conversion factors:$\Delta T_{ANT,i}=
\frac{1}{c_{i}} \Delta T_{i}$ ($c_i$ =1.29, 1.66, 3.66 ,4.82 for channels 1 to 4 respectively) and
$\rho_{i4}$ is the fraction of the astronomical signals seen at channels i and 4. The second term
within the parenthesis accounts for common structure in channels $i$ and 4, though with different
amplitudes. Based on our estimations of non-CMB contaminants, at high Galactic latitude( high $|b|$)
we expect no astronomical signal other than the CMB itself (see section~\ref{foregrounds}) so
$\rho_{i4}=1$. At low $|b|$ and in the Galactic plane we expect different signals in shape and in
amplitude as seen by different channels. This motivates the introduction of $\rho_{14} \, \neq \,
1$, whose value is obtained from our estimations of the Galactic diffuse emission.

	The atmospheric cleaning technique is applied to the 10s-binned data after subtraction of a constant offset
from each channel.  The rejected data  typically correspond  to sections taken
during bad weather and/or warming of the cryostat.  The data are divided into segments of 5
minutes. For each of these segments we compute the values of $\alpha_{i}$ from a linear fit of
channel $i$ versus channel 4. Likewise, all points within the same 5 minute segment share the same
$f_{i}$ and $f_{4}$ as computed from the splined values of water vapor w,  pressure P and temperature T from figure~\ref{fig_weather}
used as input to the code by \cite{cer85}, which computes the atmospheric opacities due to water vapor
and oxygen using the US standard atmosphere model.

\subsection{PRODUCING THE FINAL DATA SETS}

	After cleaning, a new binning is performed to bring each processed night of observation from
a 10 second to a 4 minute bin size so the beam is sampled with at least 3 points. Once again a
3$\sigma$ filter is applied to discard possible glitches occurred during the cleaning process. To
each point at 4 minutes we assign an error bar given by the standard deviation of the mean of all
points at 10s within the bin at 4 minutes. A similar analysis to that performed in
section~\ref{atmosphere}.2 allows us to obtain the mean auto-correlation function for the processed
data in each channel and both demodulations. This analysis shows that for points separated by 4
minutes the auto-correlation functions for all channels 1 to 3 and both demodulations range between
9\% (Ch 2 1F) and 1\% (Ch 1 2F). Therefore, our 4 minute bins can be considered uncorrelated as well
as their error bars . In both demodulations and in all channels we observe residual baselines of
very long periods. We proceed to remove these remnants by fitting linear combinations of sinusoidal
functions after a re-edit of the data. The re-edit discards noisy sections which may affect the
fitting process. The minimum period of the sinusoidal functions is chosen to be large enough so to
remove signals corresponding to angular scales bigger than the ones to which the instrument is
sensitive. Thus, for the 1F data the minimum period is $90\degg$ in RA and a minimum period of
$72\degg$ in RA for the 2F data. In figure~\ref{fig_process} we display these various stages of the
cleaning technique for a typical night as seen in all channels and both demodulations. In the last
column of table~\ref{tab_final_data_1} we give the mean amplitude of the baseline fits. Columns 1
and 2 show the percentage of total data used with respect to the original data at 0.25 s and the
number of nights used to generate the final data sets. The percentage of data used is bigger in
the 2F data as well as the number of used nights, with the exception of channel 3. The final
data sets were obtained by stacking all individual baseline-cleaned nights where the $rms$ did not
exceed 0.65, 1.3 and 2.5 mK for channels 1, 2 and 3 respectively in the 1F demodulation and 0.4, 0.4
and 2.0 mK for channels 1, 2 and 3 respectively in the 2F demodulation. Column 3 in
table~\ref{tab_final_data_1} gives the mean $rms$ for the surviving nights once residual baselines have been
removed. The stacking process consists of computing weighted averages, where to the $i$th bin in the
final data set we assign a value and error bar given by:

\begin{equation} 
\label{eq_stack1}
\overline{\Delta T_{i}} = \frac{\sum_{j=1}^{n_{i}} \Delta T_{ij}/\sigma_{ij}^{2}}{\sum_{j=1}^{n_{i}} 1/\sigma_{ij}^{2}}
\end{equation}

\begin{equation} 
\label{eq_stack2}
\sigma_{i}^{2} = \frac{\sum_{j=1}^{n_{i}} (\Delta T_{ij}-\overline{\Delta T_{i}})^{2}/\sigma_{ij}^{2}}
{\sum_{j=1}^{n_{i}} (n_{i}-1)/\sigma_{ij}^{2}}
\end{equation}

\noindent where the indices $ij$ refer to bin $i$ in night $j$, $\Delta T_{ij}$ and $\sigma_{ij}$ are
the data point and standard deviation at bin $i$ in night $j$, and $n_{i}$ is the number of nights
used for this $i$th bin in the final data set. The final data sets in the regions before and after
the Galactic Plane crossing are shown in figure~\ref{fig_final_sets}.

\subsection{PERFORMANCE OF THE TECHNIQUE}
 
	The efficiency of the atmospheric reduction process is best demonstrated by looking at the
power spectra of the data before and after its application. In figure~\ref{fig_ps2} we show the
power spectra of both demodulations and for channels 1, 2 and 3 for a typical night before and after
cleaning. This figure corresponds to the same data as in figure~\ref{fig_ps1}. The reduction in the
noise level is evident from these plots, so that the corrected Ch 1 and Ch 2 in both demodulations
approach the levels expected from instrument noise. We also notice the flattening, approaching the
ideal behavior of white noise. For channel 3 there is also an overall decrement in the power spectra
of both demodulations indicating that a substantial fraction of the atmospheric noise has been
subtracted. However, the levels of the cleaned data for channel 3 still show residual atmospheric
contamination. In table~\ref{tab_ps} we give the values attained by the noise spectrum in the
in-phase components at different frequencies and for both demodulations before and after cleaning
for the same observing night as in figure~\ref{fig_ps2}. The corresponding values are also given for
the out-phase component spectra. At low frequencies (i.e 0.001 Hz) where atmospheric effects are
more evident the cleaned file exhibits  lower values of the noise spectra than before cleaning; a good indication
that most of the atmospheric noise has been removed. Further, this value approaches the values
attained in the out-phase components containing small amounts of atmospheric noise due to the
leakage during the demodulation.

	Another indication of the performance of the technique is given in figure~\ref{fig_histo2}
where we show the distribution of the $rms$ values measured in the 10 second binned data after
applying our cleaning method. Once again the last bin exhibits a high value since it also contains
the contributions from all following bins to avoid a huge spread in the plots. When we compare this
figure with figure~\ref{fig_histo1} we notice a huge decrement of the $rms$ for all channels and both
demodulations.

	Another check of the performance is the recovery of the Galactic Plane (GP)
crossing. The introduction of a factor $\rho_{i4} \, \neq \, 1$ to recover the GP explicitly assumes
perfect correlation of this signal as seen by channel $i$ and channel 4. This is a very good
approximation for channels 2, 3 and 4 where the bulk of the Galactic emission is due to dust
emission. The estimated GP crossings in channel 2 1F demodulation (2F demodulation) is
 99\% (96\%) correlated with channel 4, while the estimation of the GP in channel 3 1F demodulation (2F
demodulation) attains a 100\% (100\%) correlation with respect to the GP seen by channel 4. For
channel 1 the contributions from dust and free-free emission are comparable, the latter slightly
more important. Furthermore, the free-free template (1420 MHz map) and the dust template (240 $\mu$m
DIRBE map) show a relative slight displacement in the position of the GP at $\delta=40^{\circ}$,
thus lowering the correlation between channel 1 and channel 4 down to 93\% and 72\% for the 1F and
2F data respectively. In figure~\ref{fig_gp1} we show the predicted Galactic Plane crossings
superimposed on our measurements for channels 1, 2 and 3 and both 1F and 2F demodulations. As
discussed in section~\ref{foregrounds}, we have considered the dust model in \cite{boulanger}, with
the $\rho_{i4}$ factors listed in table~\ref{tab_gp1}.  We have also tried several different
Galactic emission models in the literature and have checked that, while the absolute amplitudes in
each channel depend strongly on the model used, their ratios are quite stable and so is the parameter
$\rho_{i4}$ as can be seen in table~\ref{tab_gp1}. The slight differences between the
$\rho_{i4}$ factors for the 1F and 2F demodulations are  not relevant,  except for channel
1. This is a direct consequence of the above mentioned displacements between the contributions
generating the GP in channel 1. Using $\rho_{i4}^{\prime}$ in equation~(\ref{eq_main}) instead of
$\rho_{i4}$ results in an amplification/attenuation of the restored GP given by:

\begin{equation}
{ \Gamma_{i}} \,= \, \frac{ f_{i}/c_{i} - \alpha_{i} \cdot f_{4} /(c_{4} \cdot \rho_{i4}) }{ f_{i}/c_{i} - \alpha_{i} \cdot
f_{4} /(c_{4} \cdot \rho_{i4}^{\prime}) } 
\end{equation}

	The extreme values of $\Gamma_{i}$ obtained when using the $\rho_{i4}$ values obtained from
the  models described in table~\ref{tab_foregrounds} are 0.6 and 1.2. Therefore we do not
expect big changes in the amplitude of the restored GP due to the change of dust model used to
estimate the values of the $\rho_{i4}$ factors. The general agreement between the predictions and
our measurements constitutes an important check on the performance of our system and method.

\section{GALACTIC FOREGROUNDS}
\label{foregrounds}

	The Galactic contribution has been analyzed considering the synchrotron, free-free and dust
emission. When studying the synchrotron emission it is convenient to distinguish between the
expected levels of signal at low galactic latitudes (i.e at $|b|\stackrel{<}{_\sim} 12\degg$ or GP
crossings) and at high galactic latitudes (i.e at $|b| \stackrel{>}{_\sim} 12\degg$ or outside the
GP).

	As studied by \cite{te96}, outside the GP and at our angular scales and frequencies, the
 contributions from the Galactic synchrotron and free-free emission are expected not to be a source
 of confusion with CMB anisotropy. We have also conducted a study of the expected levels of
 contamination by synchrotron and free-free emission in our data by convolving our instrumental
 response with a template of synchrotron and free-free processes. This template was obtained by
 extrapolating in frequency from the low frequency map at 1420 MHz (\cite{reich86}), and modeling both
 processes with a single power-law: $T_{ff-sync}\propto \nu^{-\beta}$. The spectral index $\beta$ is
 obtained by fitting this law to the low frequency maps at 408 MHz (\cite{has82}) and at 1420
 MHz. The results of such an analysis predict that free-free plus synchrotron would contribute to the
 observed $rms$ with values smaller than 1~$\mu$K for both demodulations and in all
 channels. Additionally, assuming the extremely conservative approach that the signal seen at 33 GHz
 by the Tenerife experiment is entirely synchrotron emission, and extrapolating with $\beta = 2.7$, we obtain
 that for all channels the synchrotron emission at high $|b|$ amounts for less than 1$\mu K$ in both
 1F and 2F data. These two independent approaches clearly indicate the negligible effect of
 synchrotron and free-free emission on our data at high $|b|$.

	 In the GP region the map at 1420 MHz is dominated by free-free emission due to the presence
of the Cygnus X HII region and many other unresolved HII regions (\cite{Dav96}). Accordingly, when
generating the template for this region we assumed a spectral index of $\beta \simeq 2.16$ as
obtained from equation (3) of \cite{ben92}. To generate the template in this region we also
considered the possibility of still having non-negligible contributions of synchrotron emission in
the 1420 MHz map as indicated by the spectral indexes required to reproduce the GP's seen by DMR at
31.5, 53 and 90 GHz.
        
        We have used the DIRBE map at 240 $\mu m$ as template for the dust emission and extrapolated
to our frequencies by using the fit to the FIRAS data obtained by \cite{boulanger} resulting in a
model with $n=2$ and $T_{d}= 17.5 K$. We have checked the validity of this model at low $|b|$ by
reproducing the expected dust contribution to the GP seen by COBE DMR at 90 GHz where the relevant
section has been smoothed to 10\degg~FWHM. Together with a brief description and reference to the
dust model used, in table~\ref{tab_foregrounds} we give the $rms$ values expected from synchrotron
plus free-free and dust emission for different dust models outside the GP ($|b| \stackrel{>}{_\sim}
12\degg$). These $rms$ values are completely negligible as compared to the observed $rms$ values in
our final data sets (see section~\ref{statistics}). This can also be seen by using the figure of
$\Delta T_{dust}= 2.7 \pm 1.3 \mu K$ at 53 GHz and $10\degg$ angular resolution by \cite{kogut96b}.

	Of much less concern is the contamination due to unresolved sources and from known
point-like sources.  The latter contribution was computed by convolving our instrument beam with a
grid where we placed sources extracted from the K\"uhr \ea (1981) and the Green Bank sky survey
(\cite{condon}) complemented by the Michigan and Metsahovi monitoring programme.  The weakest
considered source presents a flux density at 5 GHz of 0.18 Jy. We extrapolate the flux density to
our frequencies using the fit obtained by \cite{kuhr} where available. Fluxes of sources not present
in \cite{kuhr}, but for which we have measurements at three different frequencies, were fitted to a
power law and extrapolated to our frequencies, while flat spectra were assumed for those sources for
which flux densities were available only at a single frequency. For all channels and both demodulations
the expected $rms$ in the section of our data $|b| > 12\degg$ is much smaller than 1~$\mu$K. Finally
we refer to \cite{fra89} to also exclude the contribution from randomly distributed sources given
our observing frequencies and beam width. In view of all these figures we conclude that outside the
GP the only expected astronomical signal should be CMB.

\section{STATISTICAL ANALYSIS}
\label{statistics}

\subsection{THE LIKELIHOOD FUNCTION}

	The data selected for the statistical analysis have been chosen under two
requirements. First the selected bins must correspond to regions of high $|b|$ where we are
confident that Galactic emission is negligible. Thus we have identified two sections in our data:
$RA_{1}$ for which $b \, > \, +12^{\circ}$ and $RA_{2}$ for which $b \, < \, -12^{\circ}$. The
second requirement concerns the number of independent nights ($n_{i}$ in equations~(\ref{eq_stack1})
and (\ref{eq_stack2})) used to generate the final data sets. This causes the different RA ranges
(columns 1 and 5 in table~\ref{tab_final_data_2}) over which the selected data for each channel and
demodulation span. This happens because the different editions and cleaning technique are performed
independently for each channel and demodulation. The minimum number of independent nights required is
$n_{i}$= 6, except for channels 1 and 2 in the 2F demodulation where the higher number of used nights
allows to increase this threshold up to 9 and 8 points respectively. In all cases the number of
points used for generating the final data bins is large enough to allow to use the
Kolmogorov-Smirnov test to check that our bins in the final data sets are consistent with being
drawn from Gaussian distributions. Then, the likelihood function for a set of data $\vec{x}$ is
completely specified by giving the covariance matrix $\cov$ and the vector of means $\vec{\bar{x}}$,
which in our case is identically zero. Then:

\begin{equation} 
\label{like1}
{\cal L} \, \propto \, \frac{1}{|\cov|^{\onehalf}} \exp(-\case{1}{2} \, \vec{x}^{T} \, \cov^{-1} \,\vec{x})
\end{equation}

\noindent where $|\cov|$ denotes the determinant of $\cov$. There are two independent contributions
to the covariance matrix so that $ \cov \, = \, \cov_{T} \, + \cov_{D}$ , where $\cov_{T}$
corresponds to the correlations between bins according to the model we are testing and shows
dependence on the parameters to be estimated, and $\cov_{D}$ is the data covariance matrix computed
directly from the data. We compute the elements $ij$ of $\cov_{T}$ by assuming an intrinsic Gaussian
auto-correlation function (GACF): $\cov_{intr}(\theta_{ij})= C_{0} \, \exp(-\theta_{ij}^{2}/(2\,
\theta_{c}^{2}))$ where $\theta_{ij}$ is the angular separation between bins $i$ and $j$,
$\theta_{c}$ is the coherence angle, the angle of maximum sensitivity for each experimental
configuration: $\theta_{c} \, = \, 2.17\degg$ and $\theta_{c} \, = \, 1.42\degg$ for the 1F and 2F
data respectively and $C_{0}^{1/2}$ is the parameter to be estimated. We also have to consider
the effects introduced by the finite width of the beam and the observing strategy. After convolving
with the combination of Gaussian beams reproducing our instrument responses (see
section~\ref{instrument}), we obtain that the elements $(i,j)$ of the matrices $\cov_{T,1F} $ and
$\cov_{T,2F}$, for the 1F and 2F demodulation respectively, are given by:

\begin{eqnarray}
\label{like2}
\cov_{T,1F}(i,j) & = & \aleph_{1F}^{2}\times [   2\cdot \cm (\tij;\sa,\sa,\sd) \, - \, \cm  (\tij+\beta; \sa,\sa,\sd) \, - \nonumber \\
                 &   &                          C_{M} (\tij-\beta; \sa, \sa,\sd) ]                                                    \\
                 &   &                                                                                                     \nonumber \\
\cov_{T,2F}(i,j) & = & \aleph_{2F}^{2}\times [  C_{M}(\tij;\sct,\sct,\sd) \, -\, C_{M}(\tij+\beta;\sct,\slt,\sd) \, +          \nonumber \\
                 &   &                          \frac{1}{2} C_{M}(\tij;\slt,\slt,\sd)- \,C_{M}(\tij-\beta;\sct,\slt,\sd)  \, +  \nonumber \\
                 &   &                          \frac{1}{4} ( C_{M}(\tij+2 \beta;\sct,\slt,\sd) \, + \, C_{M}(\tij+2 \beta;\sct,\slt,\sd) ) ]
\end{eqnarray}

\noindent where $ \cm (\theta;\sct,\slt,\sd) \, = \, C_{0} \theta_{c}^{2} / \sqrt{ (\theta_{c}^{2} +
\sct^{2} + \slt^{2}) (\theta_{c}^{2} + 2\sd^{2})} \, \exp [-\theta^{2}/(2 (\theta_{c}^{2} +
\sct^{2}+ \slt^{2}))]$ and, as obtained in section~\ref{instrument}, $(\sa,\sd,\beta,\aleph_{1F}) =
\, (1.^{\circ}03,0.^{\circ}86, 2.^{\circ}38,0.362)$ and $(\sct,\slt,\sd,\beta,\aleph_{2F}) = \,
(1.^{\circ}56,0.^{\circ}89,0.^{\circ}86, 2.^{\circ}40,0.593)$ for the 1F and 2F demodulation
respectively.

 In a Bayesian interpretation with uniform prior, the likelihood (${\cal L}$) as a function of the
positive definite parameter $C_{0}^{1/2}$ is directly proportional to the probability density
function of $C_{0}^{1/2}$. Then the best estimation of $C_{0}^{1/2}$ is that value for which ${\cal
L}$ is maximum while confidence levels $[(C_{0}^{1/2})_{1},(C_{0}^{1/2})_{2}]$ to a C\% level have
been computed by requiring ${\cal L}((C_{0}^{1/2})_{1}) \, = \, {\cal L}((C_{0}^{1/2})_{2})$ and
$\int_{(C_{0}^{1/2})_{1}}^{(C_{0}^{1/2})_{2}} {\cal L}(C_{0}^{1/2}) \, d(C_{0}^{1/2}) \, / \,
\int_{0}^{\infty} {\cal L}(C_{0}^{1/2}) \, d(C_{0}^{1/2}) \, = \, C/100$.  In what follows
detections are given to a 68\% CL, and upper limits to a 95\% CL . A claim of detection is made
whenever the lower limit of the confidence interval at 68\% CL is not zero, otherwise we quote the
upper limit at 95\% CL.

\subsection{LIKELIHOOD ANALYSIS ON SINGLE CHANNELS}

 	We have applied the likelihood analysis to each channel and demodulation in the ranges:
 $RA_{1}$, $RA_{2}$ and $RA_{1}$ + $RA_{2}$. The results of these analysis are given in
 table~\ref{tab_like_individual}. In all cases the values obtained in all these RA ranges are
 consistent with the presence of a common signal.  For channel 2 at 1F demodulation the results
 between $RA_{1}$ and $RA_{2}$ are still marginally consistent, but the result in the $RA_{1}$
 section shows a strong dependence on the choice of individual nights to generate the final data
 set. Accordingly in what follows we will only consider the $RA_{2}$ section for channel 2 at 1F.

	For channels 1 and 2, in all valid ranges and in both demodulations, the signals detected are
consistent between them and with values of $\sim 100 \mu$K. These values are consistent with our
previous results reported in \cite{lucio3}, where we concluded that the slight excess of signal seen
in channel 2 with respect to that obtained in channel 1 may indicate the presence of some residual
levels of atmospheric noise at the same level as the expected CMB signal. 

 	The results obtained for channel 3 at both demodulations clearly indicate that it is still
affected by important atmospheric residuals: CMB signal does not scale with frequency and our
estimated signals due to diffuse Galactic contamination at channel 3 amount to $C_{0,Gal}^{1/2} < 35
\, \mu K$,  too large a difference to be caused by the uncertainties in the Galactic estimation
procedure.

\subsection{JOINT LIKELIHOOD ANALYSIS}

	The joint analysis of all three channels allows us to estimate the most likely signal which
is common to all of  our frequencies, such as CMB anisotropy. To estimate the correlation between channels we
have computed the cross-correlation function between sections of data at 10 s which overlap in the
same nights in channels $i$ and $j$. We make use of the generalized Wiener-Khinchin relations for
stationary processes to obtain one cross-correlation curve per night. The average of these curves
are plotted in figure~\ref{fig_cross-corr}. We observe that the cross-correlation becomes negligible
at scales smaller than our binning in the final data sets, being only significantly different from
zero at zero-lags as indicated in columns 2 and 4 in table~\ref{tab_like_pair}. This
cross-correlation at zero-lag enhances the diagonal terms in the sub-matrices which take into
account the correlation between different channels in the covariance matrix $\cov_{D}$. This effect
has been analyzed and discussed in detail by \cite{gutierrez}, concluding that the net effect is an
increase of the error bars as compared with the case where not such correlations are present. In
table~\ref{tab_like_pair} we present the results from the joint analysis of any two channels and all
three channels for each demodulation . These results have been obtained by using the whole data set
except for those involving channel 2 at 1F for which only the $RA_{2}$ section of the data set was
used.

 	The analysis on any combination of 2 channels indicates the presence of common signal, which
 for the 1F demodulation is $C_{0}^{1/2} \, \sim 150\mu$K and $\sim 75\mu$K for the 2F
 demodulation. Although being completely consistent, these figures must be viewed with caution: the
 signal monotonically increases as we increase the frequencies of the channels being combined. This
 behavior again indicates the higher level of contamination in channel 3 for both demodulations.

	 Finally, we have also considered the case of having a superposition of CMB signal plus a
signal with a spectral behavior different from a black-body. In this way we obtain the contamination
due to a foreground component which is consistent with the data. We also assume a GACF with the same coherence angle as the
CMB signal for the foreground signal in addition to a scaling of the signal with frequency. Thus the
foreground signal in channel 1 exhibits an intrinsic ACF $~C^{Fgd}_{Ch~1}(\theta) = C_{0,Fgd}
\exp[-\theta^{2}/(2\theta_{c}^{2})]~$ while channel $j$ ($j=2,3$) shows $~C^{Fgd}_{Ch~j}(\theta) = C^{Fgd}_{Ch~1}(\theta) \times
(\nu_{j}/\nu_{1})^{2n} \times (c_{j}/c_{1})^{2}~$. As in equation~(\ref{eq_main})
$c_{i}~(i=1,2,3)$ is the Rayleigh-Jeans to thermodynamic conversion factor;  $\nu_{1} = 95.1$~GHz, $\nu_{2} = 169.0$~GHz
and 
$\nu_{3} = 243.5$~GHz
the frequencies for channels 1, 2 and 3 respectively. By
setting the value of $n$ to the appropriate values we obtain the signal due to any of the foreground
contaminants. The likelihood function now becomes a function of two parameters to be estimated:
${\cal L} \, = \, {\cal L}(C^{1/2}_{0,cmb},C^{1/2}_{0,Fgd})$. Since, $a~priori$, we do not have any
information about either of them we obtain the probability distribution function of $C^{1/2}_{0,cmb}$
by marginalizing with respect to $C^{1/2}_{0,Fgd}$ and vice-versa.  In table~\ref{tab_like_Fgd} we
present the results of this analysis for different values of the spectral index $n$ covering the
ranges expected for dust, atmospheric, synchrotron and free-free emission. In figures~\ref{fig_contour1} and
\ref{fig_contour2} we show the contour plots for the spectral indexes  $n \, =
$~2.0, 0.0, -2.1 and -3.0, corresponding to dust, a simple atmospheric model, free-free and synchrotron emission
 in Rayleigh-Jeans approximation. The data are then converted into thermodynamic units in the plots. Small departures from these nominal values yield essentially the same results as seen
in table 10.

\subsection{DISCUSSION OF THE RESULTS}

	The joint analysis of channels 1, 2 and 3 at the 2F demodulation reveals the presence of a
common signal with $C^{1/2}_{0} = 72^{+26}_{-24} \, \mu$K.  The conversion to band power estimates
(e.g. \cite{gacf_to_bp}) yields $ \sqrt{\lla (\lla+1) C_{\lla}/(2 \pi)} \, = \, (2.0 \pm 0.7)
\cdot 10^{-5}$ at $\lla= 53^{+22}_{-15}$ in good agreement with our previous results in
\cite{lucio3} ($C_{0}^{1/2} =76^{+23}_{-21}~\mu$K for a GACF or $ \sqrt{\lla (\lla+1)
C_{\lla}/(2\pi)} \, = \, (2.1 \pm 0.6) \cdot 10^{-5}$ in the band power estimate notation). At
similar angular scales, $\lla= 56^{+21}_{-18}$, \cite{sk95} quote a value of $ \sqrt{\lla (\lla+1)
C_{\lla}/(2 \pi)} \, = \, 1.8^{+0.7}_{-0.3} \cdot 10^{-5}$.

	We have also tested the possibility that part of the detected common signal is due to
contamination by Galactic foregrounds or residual atmospheric contamination. The analysis confirms
that dust can not be responsible of the detected signal ($C^{1/2}_{0,Dust}
< 25~\mu$K), leaving a CMB signal of $C^{1/2}_{0} =72^{+26}_{-22}~\mu$K which corresponds to
$\sqrt{\lla (\lla+1) C_{\lla}/(2\pi)} \, = \, 2.0^{+0.7}_{-0.6}\cdot 10^{-5}$. After considering calibration uncertainty as
a systematic effect the figures above become $C^{1/2}_{0} =72^{+34}_{-28}~\mu$K and 
$\sqrt{\lla (\lla+1) C_{\lla}/(2\pi)} \, = \, 2.0^{+1.0}_{-0.8}\cdot 10^{-5}$. Less conclusive is
our analysis when a free-free or synchrotron spectrum is assumed for the contaminant signal. Then we
only have  vague upper limits for the contaminants which were known in advance from both our
estimations and results from the Tenerife experiment when observing at $\delta = 40\degg$. As
 indicated in section~\ref{foregrounds}, the free-free contamination seen by the Tenerife
experiment at 33 GHz is about 4 $\mu$K. The extrapolation in frequency renders this source of
contamination to values $< 1\mu$K at both demodulations and in all channels. These upper limits are
further reduced by the fact that our experiment probes higher values of $\ell$'s than those probed
by the Tenerife experiment, and at high $|b|$ the Galactic power spectrum scales as $C_{\ell} \propto
\ell^{-3}$(\cite{te96};\cite{kogut96a}; \cite{gautier}).

	The presence of atmospheric residuals is also tested by allowing for the presence of a signal with a spectral
 index in antenna temperature of $n=0., 0.2$ and $0.4$. The $n=0$ case corresponds to the approximate case in which the
 effective atmospheric temperature is the same in all our channels. The $n=0.2$ and $0.4$ cases allow an increase with
 frequency of the effective atmospheric temperature. For all the considered $n$ values, the likelihood assigns the bulk
 of the fluctuations to CMB signal: $C^{1/2}_{0,cmb} =72^{+26}_{-24}~\mu$K, $C^{1/2}_{0,atm} <20~\mu$K. As shown in table 9, these values are rather insensitive to the exact choice of the spectral index $n$.

 	The two-component joint likelihood analysis on the 1F data places the bulk of the signal on the atmospheric
component: $C^{1/2}_{0,atm} =189^{+54}_{-39} \, \mu$K and $C^{1/2}_{0,cmb}=159^{+69}_{-63} \, \mu$K, which in flat band
power estimate becomes  $\sqrt{\lla (\lla+1) C_{\lla}/(2\pi)} \, = \,4.1^{+1.8}_{-1.6}\cdot 10^{-5}$. Treating  the
 calibration uncertainty as a systematic effect, the above results for the CMB component become
: $C^{1/2}_{0,cmb}=159^{+93}_{-87} \, \mu$K and $\sqrt{\lla (\lla+1) C_{\lla}/(2\pi)} \, = \,4.1^{+2.4}_{-2.1}\cdot 10^{-5}$. Having a large atmospheric signal in our 1F data is
expected because the 1F demodulation is known to be less efficient in removing linear gradients in the atmospheric
signal. However, a significant fraction of the signal is projected to the CMB component which is marginally consistent
 with the 2F result. When testing for the presence of Galactic contamination the two-component likelihood analyses yield
low significance detection  for a dust component, and only upper limits to contamination by free-free or synchrotron emission. 

\section{CONCLUSIONS}
\label{conclusions}

	This work demonstrates that it is possible to achieve sensitivities of a few tens of $\mu$K
in the study of CMB temperature anisotropy at millimetric wavelengths from ground based
observatories. This requires sites with stable and dry atmosphere and the combination of long
observing periods, measurements at several frequencies and a careful subtraction of the atmospheric
contribution. Other conclusions of this work are:

\begin{enumerate}

	\item The sensitivity achieved in the 2F demodulation allows us to identify a common signal
between our channels with a value of $C^{1/2}_{0} = 72^{+26}_{-24} \, \mu$K corresponding to a
band power estimate of $\sqrt{\lla (\lla+1) C_{\lla}/(2 \pi)} \, = \, 2.0 ^{+1.0}_{-0.8} \cdot
10^{-5}$ at $\lla= 53^{+22}_{-15}$ at 68\% CL including the systematic effect due to calibration uncertainty. We believe calibration
uncertainty should be treated as a systematic error and not added in quadrature to the error bar.

	\item Our value is consistent with the detection reported by the Saskatoon experiment at the
same angular scale.

	\item Two alternative arguments (one based on extrapolations from Galactic templates, the
other on results from other experiments) allow us to discard a Galactic origin of the detected
signal. An extension of the likelihood analysis allowing the presence of a signal with the assumed
spectrum for the atmospheric emission also discards the possibility of the signal being caused by
correlated atmospheric residuals.

	\item We have used a simple technique for reducing the atmospheric noise in millimetric
observations. We have presented  tests on the performance of this technique by looking at the
power spectra of the data before and after applying it. The reduction of the noise is also observed
by comparing the distribution of the noise before and after cleaning the data.

	\item We have detected the Galactic Plane crossing in all cleaned channels at both
demodulations. This constitutes the best check of the performance of the technique.

	\item In all channels the 1F data show an excess of signal with respect to the 2F data. This excess is suspected to
be of atmospheric origin. The two-component joint likelihood analysis assigns a portion of the detected signal to CMB
fluctuations at a level of $C^{1/2}_{0} = 159^{+93}_{-87} \, \mu$K or in band power estimate $\sqrt{\lla (\lla+1)
C_{\lla}/(2 \pi)} \, = \, 4.1 ^{+2.4}_{-2.1} \cdot 10^{-5}$ at 68\% CL including calibration uncertainty as a systematic effect. Our detected signal in the 1F demodulation corresponds to an
$\ell$-range never before observed.

\end{enumerate}

	We expect to improve the results presented in this work with the addition of the new
data taken during the Spring of 1996.

\acknowledgments

This work has been supported by a University of Delaware Research Foundation (UDRF) grant, by the Bartol Research Institute
and the Spanish DGICYT project PB 92-0434-c02 at the Instituto de Astrof\'{\i}sica de Canarias. We want to thank L. Page
and S. Meyer for considerable help in all the phases of this project. A special thank for the support of L. Shulman,
J. Poirer, R. Hoyland and the technical staff of the Observatorio del Teide. Special thanks to B. Keating for his careful
reading of the manuscript. Finally we would like to thank the staff from the Instituto Nacional de Meteorolog\'\i a at
Tenerife who very kindly provided us with the atmospheric data used in this analysis and to R.A. Watson for his help in the
use of these data.

\clearpage

\begin{figure}
\caption{Window function for the 1F and 2F demodulations at angular separation $\psi \, = 0$.}
\label{fig_window}
\end{figure}

\begin{figure}
\caption{Astronomical calibration by observing the Moon at the 1F demodulation. Solid lines are the
data and dashed lines the predictions using the Lunar models cited in text.}
\label{fig_moon}
\end{figure}

\begin{figure}
\caption{Meteorological conditions during the 1994 observing campaign.}
\label{fig_weather}
\end{figure}

\begin{figure}
\caption{Power spectra of the in- and out-phase components at both demodulations and for all
channels for a typical night of observation. Temperatures in this figure refer to antenna temperature values.}
\label{fig_ps1}
\end{figure}

\begin{figure}
\caption{Distribution of $rms$ values for  the whole campaign before applying the cleaning technique for
the 1F (solid line) and 2F (dash-dot line) demodulations. Note the height of the last bin in each
histogram, as it contains the contributions of that bin plus all following bins. Temperatures in
this figure refer to thermodynamic temperature values.}
\label{fig_histo1}
\end{figure}

\begin{figure}
\caption{Mean auto-correlation curves for all channels and both demodulations of the raw data. Dotted
lines represent the $\pm 1\sigma$ limits.}
\label{fig_auto-correlation}
\end{figure}

\begin{figure}
\caption{The different stages of the cleaning technique. Panels in the left column show the raw
data. Center panels display the output of equation (3), superimposed the sinusoidal fit (thick
 dashed line). Right panels show the cleaned data with the baseline removed. All plots have been
brought to a bin size of $3^{\circ}$ in RA for display purposes. Temperatures in this figure refer
to thermodynamic temperature values.}
\label{fig_process}
\end{figure}

\begin{figure}
\caption{Final data sets in the regions to be considered for the posterior Likelihood analysis. The
data have been binned to $3^{\circ}$ in RA for clarity. We show the 1F and 2F
profiles indicating the instrument response to point sources. Temperatures in this figure refer
to thermodynamic temperature values.}
\label{fig_final_sets}
\end{figure}

\begin{figure}
\caption{ Power spectra of a typical night of observation in thermodynamic temperature at 10s before  
(thin line) and after (bold line) applying the cleaning technique. The dashed lines represent the 
upper limits to the instrument noise as estimated in section 4.1. Temperatures are expressed in 
thermodynamic units.}
\label{fig_ps2}
\end{figure}

\begin{figure}
\caption{Distribution of $rms$ values after cleaning and editing for all the observing nights eligible
to be cleaned. Notice the height of the last bin in each histogram for it contains the contributions
of that bin plus all following bins. Temperatures in this figure refer to thermodynamic temperature
values.}
\label{fig_histo2}
\end{figure}

\begin{figure}
\caption{Comparison of the recovered Galactic Plane crossings at channels 1, 2 and 3 and at both
demodulations with the predictions in thermodynamic temperature.}
\label{fig_gp1}
\end{figure}

\begin{figure}
\caption{Mean cross-correlation curves between overlapping sections of individual nights used to
generate the final data sets at any two channels.}
\label{fig_cross-corr}
\end{figure}

\begin{figure}
\caption{Contour plots of the likelihood surface of the joint analysis on channels 1,2 and 3 for the 1F
demodulation when a second component other than CMB is allowed. The four indeces represent the four
relevant foregrounds: dust, atmosphere,free-free and synchrotron emission. Contour levels represent
the confidence levels at 68\% (solid line) and 95\% (dashed line). The X symbol indicates the position of  the likelihood
surface peak. Temperatures on the axis are in thermodynamic units.}
\label{fig_contour1}
\end{figure}

\begin{figure}
\caption{Contour plots of the likelihood surface of the joint analysis on channels 1,2 and 3 for the 1F
demodulation when a second component other than CMB is allowed. The four indeces represent the four
relevant foregrounds: dust, atmosphere,free-free and synchrotron emission. Contour levels represent
the confidence levels at 68\% (solid line) and 95\% (dashed line). The X symbol indicates the position of  the likelihood
surface peak. Temperatures on the axis are in thermodynamic units.}
\label{fig_contour2}
\end{figure}


\clearpage 
\begin{deluxetable}{lccccc}
\footnotesize
\tablewidth{456.150960pt}
\tablecaption{Results From Fits To Extended Moon Transits.}
\tablehead{ \colhead{} & \multicolumn{2}{c}{1F DEMODULATION} & \colhead{} & \multicolumn{2}{c}{2F DEMODULATION} \\
\cline{2-3} \cline{5-6} \\
\colhead{} & \colhead{$\sigma$} &  \colhead{$\beta_{0}$} & \colhead{} &\colhead{$\sigma$} & \colhead{$\beta_{0}$} }
\startdata
     Ch 1  &  $0.94 \pm 0.10$   & $2.89 \pm 0.09$       &    &  $1.01 \pm 0.27$  & $2.80 \pm 0.29$    \\  
     Ch 2  &  $0.88 \pm 0.07$   & $2.90 \pm 0.06$       &    &  $0.92 \pm 0.19$  & $2.85 \pm 0.19$    \\   
     Ch 3  &  $0.85 \pm 0.06$   & $2.90 \pm 0.05$       &    &  $0.88 \pm 0.17$  & $2.86 \pm 0.16$    \\   
     Ch 4  &  $0.82 \pm 0.07$   & $2.91 \pm 0.06$       &    &  $0.84 \pm 0.13$  & $2.88 \pm 0.13$    \\   
\enddata
\label{tab_fits_moon}
\end{deluxetable}


\clearpage
\begin{deluxetable}{lccccc}
\footnotesize
\tablewidth{456.150960pt}
\tablecaption{Correlation Between Channels Before Applying Atmospheric Correction.}
\tablehead{ \colhead{} & \multicolumn{2}{c}{1F DEMODULATION} & \colhead{} & \multicolumn{2}{c}{2F DEMODULATION} \\
\cline{2-3} \cline{5-6} \\
\colhead{} & \colhead{Whole Campaign} & \colhead{Final Data Set} & \colhead{} & \colhead{Whole Campaign} &
 \colhead{Final Data Set} }	
\startdata
Ch 1 - Ch 4&  0.965   $\pm$ 0.010  & 0.990   $\pm$ 0.010  &   &  0.884  $\pm$ 0.023  &   0.91  $\pm$ 0.09  \phn \nl	 
Ch 2 - Ch 4&  0.983   $\pm$ 0.008  & 0.995   $\pm$ 0.011  &   &  0.977  $\pm$ 0.007  &   0.991 $\pm$ 0.010 \phn \nl
Ch 3 - Ch 4&  0.99963 $\pm$ 0.00010& 0.99984 $\pm$ 0.00010&   &  0.9956 $\pm$ 0.0011 &   0.995 $\pm$ 0.010 \phn \nl
\enddata
\label{tab_correlations1}
\end{deluxetable}


\clearpage
\begin{deluxetable}{lcccc}
\footnotesize
\tablewidth{456.150960pt}
\tablecaption{Data Used To Generate Final Data Sets.}
\tablehead{
 \multicolumn{5}{c}{1F DEMODULATION}  \\
\cline{1-5}  \\
\colhead{} & \colhead{Percentage} & \colhead{Number nights} & \colhead{Mean $rms$ (mK)}&\colhead{Amplitude Baseline (mK)} }
\startdata
Ch 1 &	17.2 \% & 16 & 0.392 $\pm$ 0.024  & 0.78 $\pm$ 0.19   \phn \nl
Ch 2 &  13.3 \% & 14 & 0.48  $\pm$ 0.07   & 1.23 $\pm$ 0.23   \phn \nl
Ch 3 &  16.8 \% & 12 & 1.30  $\pm$ 0.14   & 1.81 $\pm$ 0.23   \phn \nl
\cutinhead{2F DEMODULATION}
Ch 1 &  30.1 \% & 24 & 0.275 $\pm$ 0.024  & 0.46 $\pm$ 0.08   \phn \nl
Ch 2 &  15.4 \% & 18 & 0.233 $\pm$ 0.018  & 0.85 $\pm$ 0.23   \phn \nl
Ch 3 &  14.5 \% & 15 & 0.86  $\pm$ 0.12   & 1.28 $\pm$ 0.25   \phn \nl
\enddata
\tablecomments{Mean $rms$ and baseline amplitudes are expressed in thermodynamic units.}
\label{tab_final_data_1}
\end{deluxetable}


\clearpage
\begin{deluxetable}{ccccccccc}
\scriptsize
\tablewidth{456.15096pt}
\tablecaption{Noise Spectrum ($mK~s^{1/2}$) Before And After Applying Atmospheric Subtraction.}
\tablehead{ \colhead{} & \colhead{1F Out} & \colhead{1F Out} & \colhead{1F In} & \colhead{1F In}
                       & \colhead{2F Out} & \colhead{2F Out} & \colhead{2F In} & \colhead{2F In} \\
     \colhead{Channel} & \colhead{1 Hz}         & \colhead{0.001 Hz}     & \colhead{0.001 Hz}    & \colhead{0.001 Hz} 
                       & \colhead{1 Hz}         & \colhead{0.001 Hz}     & \colhead{0.001 Hz}    & \colhead{0.001 Hz}    \\
            \colhead{} & \colhead{}             & \colhead{}             &  \colhead{}           & \colhead{CLEANED} 
                       & \colhead{}             & \colhead{}             &  \colhead{}           & \colhead{CLEANED}         }
\startdata
1 &  4.8$\pm$0.4  &  6.9$\pm$0.6  &  52$\pm$5  &  8.1$\pm$1.0 &  3.6$\pm$0.3  & 6.7$\pm$0.6 &  8.8$\pm$0.7 &  6.1$\pm$0.7 \phn \nl
2 & 1.13$\pm$0.10 & 1.87$\pm$0.17 & 115$\pm$10 &  3.9$\pm$0.5 & 1.04$\pm$0.11 & 3.5$\pm$0.4 & 12.3$\pm$1.1 &  3.3$\pm$0.4 \phn \nl
3 & 1.99$\pm$0.18 &  3.6$\pm$0.3  & 570$\pm$50 & 20.0$\pm$2.3 & 1.81$\pm$0.17 &16.0$\pm$1.5 &   55$\pm$5   & 12.7$\pm$1.5 \phn \nl
4 & 1.20$\pm$0.11 & 18.9$\pm$1.7  & 750$\pm$60 &    \nodata   & 1.07$\pm$0.10 &27.6$\pm$2.6 &   94$\pm$8   &      \nodata \phn \nl
\enddata
\tablecomments{Mean $rms$ and baseline amplitudes are expressed in thermodynamic units.}
\label{tab_ps}
\end{deluxetable}


\clearpage
\begin{deluxetable}{ccccccccc}
\scriptsize
\tablewidth{456.150960pt}
\tablecaption{Values Of $\rho_{i4}$ For The Models Under Consideration.}
\tablehead{ \multicolumn{2}{c}{ }& \multicolumn{3}{c}{1F DEMODULATION} & \colhead{} & \multicolumn{3}{c}{2F DEMODULATION} 
\\ \cline{3-5} \cline{7-9} \\
\colhead{Model} & \colhead{Reference}  &  \colhead{$\rho_{14}$} &  \colhead{$\rho_{24}$} &  \colhead{$\rho_{34}$} &
\colhead{}      &                         \colhead{$\rho_{14}$} &  \colhead{$\rho_{24}$} &  \colhead{$\rho_{34}$} }
\startdata
 1  & 1 & 0.1167  & 0.1515  & 0.6458  &    & 0.1567  & 0.1533  & 0.6452     \phn \nl
 2  & 1 & 0.1249  & 0.1533  & 0.6453  &    & 0.1750  & 0.1551  & 0.6446     \phn \nl
 3  & 3 & 0.1292  & 0.1786  & 0.6724  &    & 0.1458  & 0.1787  & 0.6720     \phn \nl
 4  & 3 & 0.1291  & 0.1720  & 0.6624  &    & 0.1505  & 0.1720  & 0.6618     \phn \nl
 5  & 4 & 0.1233  & 0.1528  & 0.6454  &    & 0.1725  & 0.1546  & 0.6448     \phn \nl
 6  & 5 & 0.1229  & 0.1859  & 0.6778  &    & 0.1273  & 0.1851  & 0.6775     \phn \nl
 7  & 6 & 0.1147  & 0.1555  & 0.6511  &    & 0.1397  & 0.1556  & 0.6506     \phn \nl
 8  & 7 & 0.1137  & 0.1780  & 0.6739  &    & 0.1178  & 0.1773  & 0.6736     \phn \nl
 9  & 8 & 0.1195  & 0.1788  & 0.6723  &    & 0.1269  & 0.1779  & 0.6719     \phn \nl
 10 & 9 & 0.1386  & 0.2229  & 0.7063  &    & 0.1379  & 0.2224  & 0.7061     \phn \nl
 11 &10 & 0.1552  & 0.1654  & 0.6438  &    & 0.1678  & 0.2105  & 0.6427     \phn \nl
 12 &11 & 0.1299  & 0.1748  & 0.6649  &    & 0.1464  & 0.1749  & 0.6644     \phn \nl
\enddata
\tablerefs{
(1) Boulanger \etal 1996; (2) Reach \etal 1995; (3) Wright \etal 1991; (4) Bersanelli \etal 1995; (5) Davies et
al. 1996a;(6) Kogut \etal 1996a; (7) Kogut \etal 1996b; (8) de Bernardis \etal 1991; (9) Banday \& Wolfendale 1991; (10)
Page \etal 1990; (11) Fischer \etal 1995. } 
\label{tab_gp1}
\end{deluxetable}


\clearpage
\begin{deluxetable}{ccccccccc}
\scriptsize
\tablewidth{456.150960pt}
\tablecaption{Expected $rms$ Values Due To Diffuse Galactic Emission. Units Of $\mu K$.}
\tablehead{ \multicolumn{3}{c}{ }& \multicolumn{3}{c}{1F DEMODULATION} & \multicolumn{3}{c}{2F DEMODULATION} 
 \\ \colhead{Model} & \colhead{Reference}  & \colhead{Description} &
\colhead{$rms_{Ch1}$}  & \colhead{$rms_{Ch2}$}  &\colhead{$rms_{Ch3}$} &  
\colhead{$rms_{Ch1}$}  & \colhead{$rms_{Ch2}$}  &\colhead{$rms_{Ch3}$} }
\startdata
 1  & 1 & $I_{\nu}^{D} \propto \nu^{2} B_{\nu}(17.5)$    & 1.2 &  2.4  & 12.3   & 0.6 & 1.3 &  6.7  \phn \nl
 2  & 1 & $I_{\nu}^{D} \propto \nu^{2} B_{\nu}(18.2)$    & 1.1 &  2.2  & 11.3   & 0.6 & 1.2 &  6.1  \phn \nl
 3  & 2 & $I_{\nu}^{D} \propto \nu^{2} (B_{\nu}(20.4)+$  & 1.5 &  3.1  & 13.6   & 0.8 & 1.7 &  7.4  \phn \nl
    &   & $~~~~~~~~~+~6.7 \times B_{\nu}(4.8) )$         &     &       &        &     &     &       \phn \nl
 4  & 3 & $I_{\nu}^{D} \propto \nu^{1.65} B_{\nu}(23.2)$ & 1.5 &  2.9  & 13.1   & 0.8 & 1.6 &  7.1  \phn \nl
 5  & 4 & $I_{\nu}^{D} \propto \nu^{2} B_{\nu}(18.0)$    & 1.1 &  2.3  & 11.6   & 0.6 & 1.2 &  6.3  \phn \nl
 6  & 5 & $I_{\nu}^{D} \propto \nu^{1.4} B_{\nu}(23.3)$  & 2.5 &  4.9  & 19.6   & 1.3 & 2.6 & 10.6  \phn \nl
 7  & 6 & $I_{\nu}^{D} \propto \nu^{1.9} B_{\nu}(18)$    & 1.3 &  2.8  & 13.7   & 0.7 & 1.5 &  7.4  \phn \nl
 8  & 7 & $I_{\nu}^{D} \propto \nu^{1.5} B_{\nu}(20)$    & 2.5 &  5.2  & 21.4   & 1.3 & 2.8 & 11.6  \phn \nl
 9  & 8 & $I_{\nu}^{D} \propto \nu^{1.5} B_{\nu}(22)$    & 2.2 &  4.3  & 18.2   & 1.1 & 2.3 &  9.8  \phn \nl
 10 & 9 & $I_{\nu}^{D} \propto \nu B_{\nu}(22.1)$        & 7.0 & 12.5  & 41.0   & 3.8 & 6.8 & 22.2  \phn \nl
 11 &10 & $I_{\nu}^{D} \propto \nu^{2} B_{\nu}(22.1)$    & 0.9 &  1.5  &  8.0   & 0.5 & 0.8 &  4.3  \phn \nl
 12 &11 & $I_{\nu}^{D} \propto \nu^{1.6} B_{\nu}(24)$    & 1.6 &  3.1  & 13.5   & 0.8 & 1.6 &  7.3  \phn \nl
\enddata
\tablerefs{
(1) Boulanger \etal 1996; (2) Reach \etal 1995; (3) Wright \etal 1991; (4) Bersanelli \etal 1995; (5) Davies et
al. 1996a;(6) Kogut \etal 1996a; (7) Kogut \etal 1996b; (8) de Bernardis \etal 1991; (9) Banday \& Wolfendale 1991; (10)
Page \etal 1990; (11) Fischer \etal 1995.}
\tablecomments{$I_{\nu}^{D}$ and $B_{\nu}(T)$ stand for the dust spectrum and a black-body spectrum at a temperature of $T$
 Kelvin, respectively. Models (3) and (4) differ in the dust spectrum at low $|b|$. }
\label{tab_foregrounds}
\end{deluxetable}


\clearpage
\begin{deluxetable}{llcccclccc}
\scriptsize
\tablewidth{456.150960pt}
\tablecaption{Basic Statistic Figures Of Final Data Sets.}
\tablehead{
\colhead{} & \multicolumn{4}{c}{$RA_{1}$} & \colhead{} & \multicolumn{4}{c}{$RA_{2}$}  \\
\cline{2-5} \cline{7-10} \\
\colhead{} & \colhead{Range}     & \colhead{Mean}      & \colhead{Mean}             & \colhead{Mean} & \colhead{}
           & \colhead{Range}     & \colhead{Mean}      & \colhead{Mean}             & \colhead{Mean}  \\ 
\colhead{} & \colhead{($\degg$)} & \colhead{\# points} & \colhead{$\sigma~(\mu K)$} & \colhead{$rms~(\mu K)$} & \colhead{}
           & \colhead{($\degg$)} & \colhead{\# points} & \colhead{$\sigma~(\mu K)$} & \colhead{$rms~(\mu K)$} }
\startdata
Ch 1 1F  & [224,285] &  7.9 & 130.6 & 134.3  &   &   [331,369] & 10.8 & 103.1 & 123.7 \phn \nl
Ch 2 1F  & [230,285] &  8.3 & 118.9 & 157.9  &   &   [331,361] &  7.4 & 113.2 & 110.6 \phn \nl
Ch 3 1F  & [224,285] &  8.8 & 312.6 & 340.5  &   &   [331,367] &  8.7 & 310.6 & 443.6 \phn \nl
         &           &      &       &        &   &             &      &       &       \phn \nl
Ch 1 2F  & [206,285] & 12.6 &  74.7 &  62.5  &   &   [331,381] & 13.2 &  74.1 &  74.4 \phn \nl
Ch 2 2F  & [229,285] &  9.1 &  68.5 &  71.0  &   &   [331,360] &  9.5 &  53.3 &  58.5 \phn \nl
Ch 3 2F  & [236,285] &  8.3 & 236.4 & 197.9  &   &   [331,362] &  9.5 & 187.7 & 204.7 \phn \nl
\enddata
\tablecomments{Mean $\sigma$ refers to the mean error bar associated with each in the final data set. Mean $rms$ is the
weighted $rms$ along the indicated RA range. Both quantities expressed in thermodynamic units.}
\label{tab_final_data_2}
\end{deluxetable}

\clearpage
\begin{deluxetable}{lccccccc}
\footnotesize
\tablewidth{456.150960pt}
\tablecaption{Likelihood Results On Individual Data Sets. Values In $\mu$K CMB.}
\tablehead{ \colhead{} & \multicolumn{3}{c}{1F DEMODULATION} & \colhead{} & \multicolumn{3}{c}{2F DEMODULATION} \\
\cline{2-4} \cline{6-8} \\
\colhead{Channel} & \colhead{$(C_{0})^{1/2}$} & \colhead{$(C_{0})^{1/2}$} & \colhead{$(C_{0})^{1/2}$} & \colhead{} & \colhead{$(C_{0})^{1/2}$} & \colhead{$(C_{0})^{1/2}$}& \colhead{$(C_{0})^{1/2}$} \\
 \colhead{}& \colhead{$RA_{1}$} & \colhead{$RA_{2}$} & \colhead{$RA_{1}$+$RA_{2}$} &
\colhead{} & \colhead{$RA_{1}$} & \colhead{$RA_{2}$} & \colhead{$RA_{1}$+$RA_{2}$}  }
\startdata
Ch 1 & $130^{+ 63}_{- 52}$ & $124^{+ 71}_{- 58}$ & $127^{+ 44}_{- 37}$ &
     & $  < \, 123       $ & $ 99^{+ 53}_{- 45}$ & $ 71^{+ 34}_{- 37}$ \phn \nl
Ch 2 & \nodata             & $155^{+ 83}_{- 58}$ & $155^{+ 83}_{- 58}$  &  
     & $ 91^{+ 47}_{- 43}$ & $132^{+ 77}_{- 56}$ & $106^{+ 37}_{- 33}$ \phn \nl
Ch 3 & $764^{+219}_{-169}$ & $626^{+260}_{-176}$ & $711^{+168}_{-135}$ &  
     & $235^{+158}_{-181}$ & $591^{+284}_{-240}$ & $373^{+169}_{-163}$ \phn \nl
\enddata
\tablecomments{The stated confidence intervals do not include calibration uncertainties.}
\label{tab_like_individual}
\end{deluxetable}


\clearpage
\begin{deluxetable}{cccccc}
\footnotesize
\tablewidth{456.15096pt}
\tablecaption{Joint Likelihood Results. Values In $\mu$K CMB.}
\tablehead{ \colhead{} & \multicolumn{2}{c}{1F DEMODULATION} & \colhead{} & \multicolumn{2}{c}{2F DEMODULATION} \\
\cline{2-3} \cline{5-6} \\
\colhead{Channels} & \colhead{$(C_{0})^{1/2}$ ($\mu$K)} & \colhead{Cross-Corr $i$-$j$} & \colhead{} &
\colhead{$(C_{0})^{1/2}$ ($\mu$K)} & \colhead{Cross-Corr $i$-$j$} }
\startdata
Ch 1 x Ch 2       &$129^{+41}_{-35}$& $0.19  \pm 0.07$  &  &$ 69^{+27}_{-25}$ & $0.149 \pm 0.026$ \phn \nl 
Ch 1 x Ch 3       &$151^{+44}_{-37}$& $0.085 \pm 0.013$ &  &$ 72^{+35}_{-37}$ & $0.049 \pm 0.010$ \phn \nl 
Ch 2 x Ch 3       &$164^{+89}_{-60}$& $0.31  \pm 0.06$  &  &$107^{+37}_{-34}$ & $0.26  \pm 0.05 $ \phn \nl 
Ch 1 x Ch 2 x Ch 3&$150^{+40}_{-34}$&      \nodata      &  &$ 72^{+26}_{-24}$ &       \nodata     \phn \nl
\enddata
\tablecomments{The stated confidence intervals do not include calibration uncertainties.}
\label{tab_like_pair}
\end{deluxetable}


\clearpage
\begin{deluxetable}{lcccccc}
\footnotesize
\tablewidth{456.15096pt}
\tablecaption{Likelihood Results Allowing A Foreground Component. Values In $\mu$K CMB.}
\tablehead{
\colhead{} & \colhead{} & \multicolumn{2}{c}{1F DEMODULATION} & \colhead{} & \multicolumn{2}{c}{2F DEMODULATION}\\
  \\ \cline{3-4} \cline{6-7}  \\
\colhead{n} & \colhead{Foreground} & \colhead{$C^{1/2}_{0,cmb}~(\mu K)$} & \colhead{$C^{1/2}_{0,Fgd}~(\mu K)$} & \colhead{} &
 \colhead{$C^{1/2}_{0,cmb}~(\mu K)$} & \colhead{$C^{1/2}_{0,Fgd}~(\mu K)$} }
\startdata
+2.0 & DUST        & $128^{+42}_{-32}$ & $ 38^{+ 8}_{- 6}$ & & $72^{+26}_{-22}$ & $< 6 $ \phn \nl
+1.5 & DUST        & $128^{+42}_{-34}$ & $ 60^{+14}_{-10}$ & & $72^{+26}_{-22}$ & $< 8 $  \phn \nl
+0.4 & ATMOSPHERE  & $147^{+54}_{-38}$ & $150^{+39}_{-27}$ & & $72^{+26}_{-22}$ & $< 16$  \phn \nl
+0.2 & ATMOSPHERE  & $154^{+62}_{-54}$ & $171^{+45}_{-33}$ & & $72^{+26}_{-24}$ & $< 18$  \phn \nl
+0.0 & ATMOSPHERE  & $159^{+69}_{-63}$ & $189^{+54}_{-39}$ & & $72^{+26}_{-24}$ & $< 20$  \phn \nl
-1.8 & FREE-FREE   & $147^{+45}_{-39}$ &    $< 177$        & & $72^{+26}_{-26}$ & $< 62$  \phn \nl
-2.1 & FREE-FREE   & $150^{+45}_{-39}$ &    $< 189$        & & $72^{+26}_{-26}$ & $< 74$  \phn \nl
-2.4 & FREE-FREE   & $153^{+51}_{-39}$ &    $< 201$        & & $72^{+26}_{-26}$ & $< 86$  \phn \nl
-2.7 & SYNCHROTRON & $153^{+51}_{-39}$ &    $< 210$        & & $72^{+26}_{-26}$ & $< 96$  \phn \nl
-3.0 & SYNCHROTRON & $156^{+51}_{-39}$ &    $< 216$        & & $72^{+26}_{-28}$ & $<106$  \phn \nl
-3.3 & SYNCHROTRON & $156^{+51}_{-39}$ &    $< 222$        & & $72^{+26}_{-28}$ & $<116$  \phn \nl
\enddata
\label{tab_like_Fgd}
\tablecomments{The stated confidence intervals do not include calibration uncertainties.}
\end{deluxetable}

\clearpage

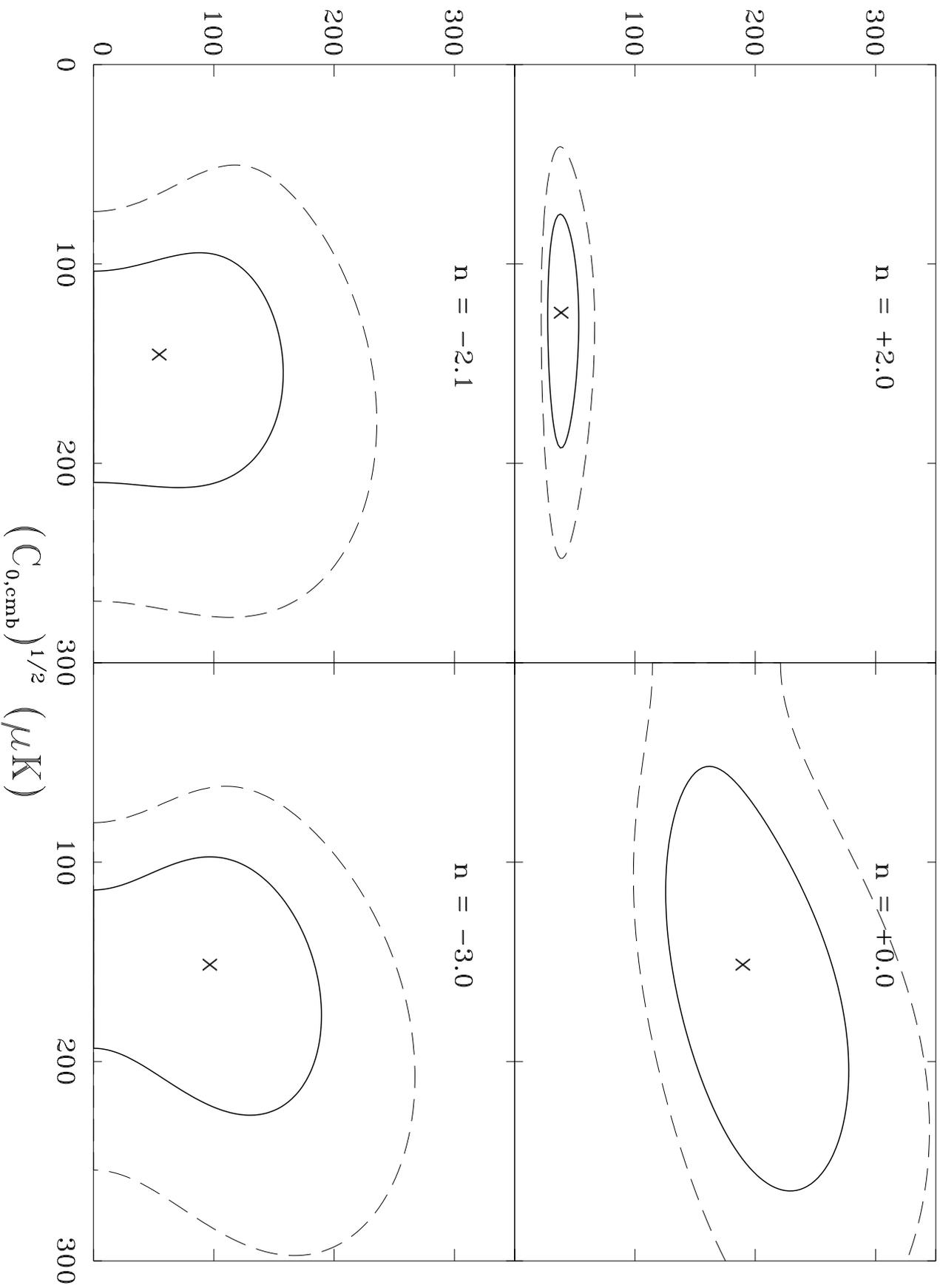
\end{document}